\documentclass[aps,reprint,showpacs,floatfix]{revtex4-1}
\usepackage{amsmath,graphicx,color,verbatim}

\begin{document}

\title{Performance of the strongly constrained and appropriately normed density functional for solid-state materials}
\author{Eric B. Isaacs}
\author{Chris Wolverton}
\email{c-wolverton@northwestern.edu}
\affiliation{Department of Materials Science and Engineering, Northwestern University, Evanston, Illinois 60208, USA}

\begin{abstract}
Constructed to satisfy all known exact constraints and appropriate
norms for a semilocal density functional, the strongly constrained and
appropriately normed (SCAN) meta-generalized gradient approximation
functional has shown early promise for accurately describing the
electronic structure of molecules and solids. One open question is how
well SCAN predicts the formation energy, a key quantity for describing
the thermodynamic stability of solid-state compounds. To answer this
question, we perform an extensive benchmark of SCAN by computing the
formation energies for a diverse group of nearly one thousand
crystalline compounds for which experimental values are known. Due to
an enhanced exchange interaction in the covalent bonding regime, SCAN
substantially decreases the formation energy errors for strongly-bound
compounds, by approximately 50\% to 110 meV/atom, as compared to the
generalized gradient approximation of Perdew, Burke, and Ernzerhof
(PBE). However, for intermetallic compounds, SCAN performs moderately
worse than PBE with an increase in formation energy error of
approximately 20\%, stemming from SCAN's distinct behavior in the weak
bonding regime. The formation energy errors can be further reduced via
elemental chemical potential fitting. We find that SCAN leads to
significantly more accurate predicted crystal volumes, moderately
enhanced magnetism, and mildly improved band gaps as compared to PBE.
Overall, SCAN represents a significant improvement in accurately
describing the thermodynamics of strongly-bound compounds.
\end{abstract}

\date{\today}
\maketitle

\section{Introduction}\label{sec:intro}

Density functional theory (DFT)
\cite{hohenberg_inhomogeneous_1964,kohn_self-consistent_1965} is the
standard approach for computing the electronic structure of
solid-state materials due to an attractive balance between accuracy
and computational efficiency in the Kohn-Sham approach. If the
underlying exchange-correlation (xc) energy functional $E_{xc}$ were
known, in principle DFT would give the exact ground-state properties
of any many-electron system. In practice, the exact $E_{xc}$ is
unknown and must be approximated. Examples of such approximations are
the well-known local density approximation (LDA) and generalized
gradient approximation (GGA). Developing improved approximate
functionals for DFT is an important challenge for electronic structure
theory.

\begin{figure}[htbp]
  \begin{center}
    \includegraphics[width=1.0\linewidth]{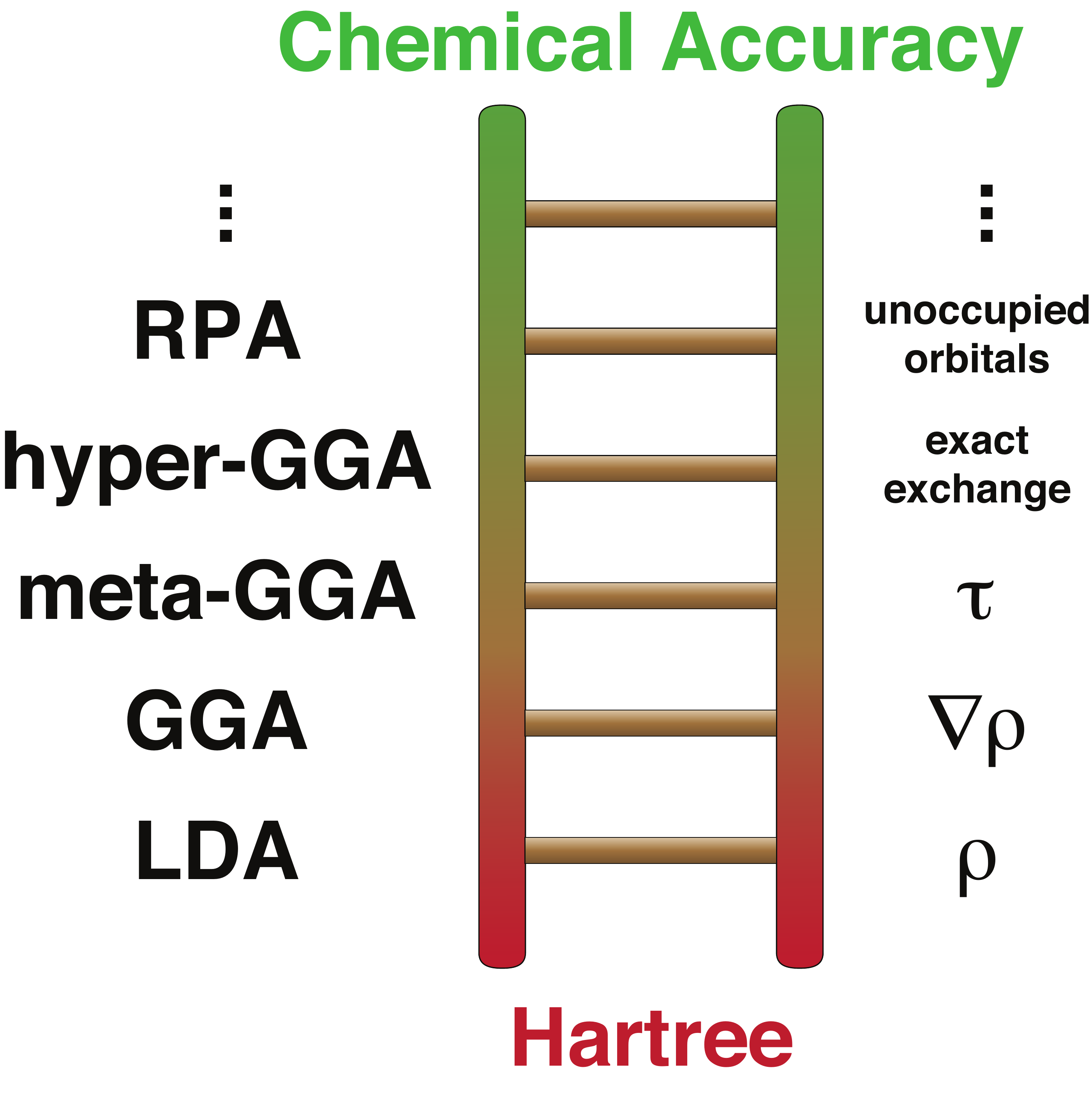}
  \end{center}
  \caption{Jacob's ladder framework for describing different levels of
    sophistication and accuracy of $E_{xc}$ in density functional
    theory. Starting from the Hartree level of theory ($E_{xc}=0$) and
    attempting to climb towards chemical accuracy, each rung of the
    ladder labeled on the left corresponds to additional dependencies
    of $E_{xc}$ indicated on the right. The top rung listed is the
    random phase approximation (RPA). Figure is based on Refs.
    \citenum{perdew_jacobs_2001} and
    \citenum{perdew_prescription_2005}.}
  \label{fig:jacob_ladder}
\end{figure}

A substantial weakness of DFT is that, unlike certain quantum
chemistry approaches, there is no straightforward way to
systematically converge to the exact result. In other words, it is not
generally clear how to develop increasingly more accurate
approximations to $E_{xc}$. However, Perdew has proposed a general
framework to describe and develop improvements to $E_{xc}$ by
including dependence on additional information. In this framework,
known as Jacob's ladder \cite{perdew_jacobs_2001}, each rung of the
ladder is a more sophisticated approximation to $E_{xc}$, as shown in
Fig. \ref{fig:jacob_ladder}. In the bottom rung, LDA, $E_{xc}$ depends
solely the electron density $\rho$. The next rung up is GGA, in which
$E_{xc}$ depends on $\nabla\rho$ in addition to $\rho$ itself (for
example, in the GGA of Perdew, Burke, and Ernzerhof (PBE)
\cite{perdew_generalized_1996}). Further up the ladder are more
complex $E_{xc}$ functionals containing explicit dependence on
Kohn-Sham wavefunctions $\psi$, such as hybrid functionals. While
beyond-DFT approaches like DFT+$U$
\cite{himmetoglu_hubbard-corrected_2014} and DFT plus dynamical
mean-field theory (DMFT) \cite{kotliar_electronic_2006} have not been
considered in the Jacob's ladder framework, they share a similar
philosophy. In these methodologies, the energy functional relies on
the local density matrix or local Green function, for a set of
localized orbitals, in addition to $\rho$, $\nabla\rho$, etc.

Meta-generalized gradient approximation (meta-GGA), the third rung of
Jacob's ladder, takes the $E_{xc}$ form of GGA and adds an additional
dependence on the positive orbital kinetic energy density
\begin{equation}\tau = \sum_i \frac{1}{2} |\nabla\psi_i|^2.\end{equation} Here $\psi_i$
is the Kohn-Sham wavefunction for the $i$th occupied band. $E_{xc}$
now depends on $\tau$ in addition to $\rho$ and $\nabla\rho$:
  \begin{equation}\label{eq:mgga_exc}
    E_{xc}[\rho, \nabla\rho, \tau] = \int \rho
    \epsilon_{xc}(\rho,\nabla\rho,\tau)\ d^3r
  \end{equation}
Here we consider a spin-dependent meta-GGA, just as the local spin
density approximation (LSDA) is the spin-dependent version of LDA.
Therefore, just as $E_{xc}$ of LSDA depends on the different spin
densities ($\rho_{\uparrow}$ and $\rho_{\downarrow}$) separately, here
$E_{xc}$ depends on the different $\rho$, $\nabla\rho$, and $\tau$ for
each spin channel. For brevity, we do not indicate the separate spin
channels in Eq. \ref{eq:mgga_exc}.

We note that, more generally, the meta-GGA rung of Jacob's ladder also
includes $E_{xc}$ that depends on $\nabla^2\rho$ in addition to
$\tau$. There is evidence, however, that $\tau$ contains essentially
the same information as $\nabla^2\rho$
\cite{perdew_laplacian-level_2007}. Therefore, here we only comment on
meta-GGA depending solely on $\tau$. $\tau$ is an implicit functional
of $\rho$ via the Hohenberg-Kohn theorem. Therefore, meta-GGA can
still be considered a pure density functional. Meta-GGA functionals
are nonlocal since $\tau$ is not local in $\rho$. However, they can
still be considered as semilocal DFT since they are not explicitly
nonlocal in $\rho$. Meta-GGA are semilocal in $\psi$. Although
nonlocal density functionals (involving a double integral in $\rho$)
can be much more computationally intensive, this is not so in the case
of meta-GGA since the nonlocality stems only from the dependence on
$\psi$, which is readily available.

In 2015, Sun \textit{et al.} introduced the strongly constrained and
appropriately normed (SCAN) functional, a new non-empirical meta-GGA
functional \cite{sun_strongly_2015}. SCAN satisfies all known possible
exact constraints for a meta-GGA functional. One example is the
requirement that the exchange enhancement factor $F_x = E_x/E_x^{LDA}$
must be no larger than 1.174, a constraint derived from the case of a
non-spin-polarized 2-electron density \cite{perdew_gedanken_2014}.
Here $E_x$ is the exchange part of $E_{xc}$ ($E_c$ is the correlation
part) and $E_x^{LDA}$ is the LDA $E_x$. SCAN is also designed to
accurately describe particular systems for which exact results are
known, which are known as norms. The simplest example of such a norm
is the homogeneous electron gas (jellium), which is exactly described
by LDA by construction. Examples of norms for SCAN include the jellium
surface, as well as the large-$Z$ scaling behavior of the $E_x$ and
$E_c$ for noble gas atoms, where $Z$ is the atomic number. The norms
chosen are called \textit{appropriate} in the sense that a meta-GGA
should in principle be able to describe them. Stretched H$_2^+$ is an
example of an inappropriate norm for meta-GGA since in this case the
xc hole will be far from the reference electron
\cite{perdew_semilocal_2016}.

The SCAN functional depends on a dimensionless measure of $\tau$
called $\alpha$ defined
as: \begin{equation}\label{eq:alpha}\alpha=\frac{\tau-\tau_{\textrm{single-orbital}}}{\tau_{\textrm{uniform}}}\end{equation}
Here $\tau_{\textrm{single-orbital}} = |\nabla \rho|^2/8\rho$ and
$\tau_{\textrm{uniform}} = (3/10)(3\pi^2)^{2/3}\rho^{5/3}$ are the
limits of $\tau$ for the single orbital and uniform density cases,
respectively. The electron localization function
\cite{becke_simple_1990,silvi_classification_1994} can be written very
simply in terms of $\alpha$ as $(1+\alpha^2)^{-1}.$

\begin{figure}[htbp]
  \begin{center}
    \includegraphics[width=1.0\linewidth]{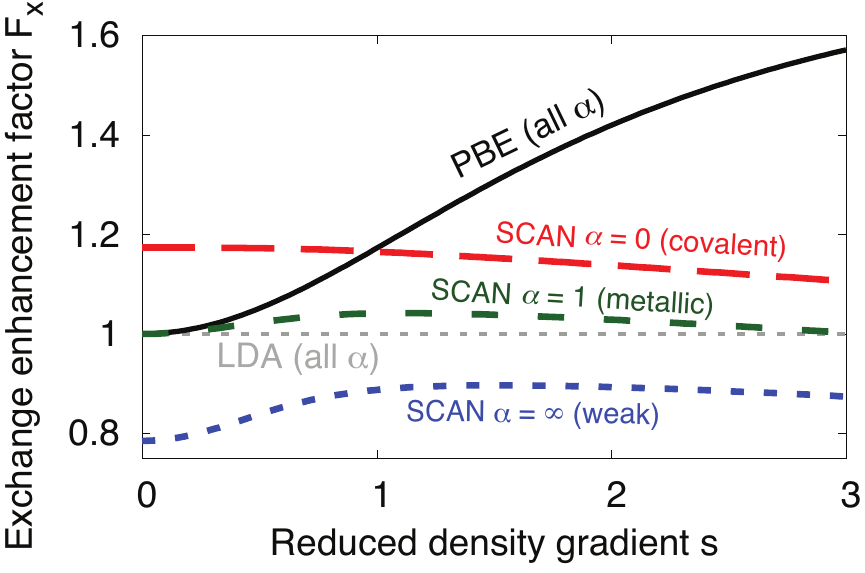}
  \end{center}
  \caption{Non-spin-polarized exchange enhancement factor for the LDA,
    PBE, and SCAN density functionals as a function of dimensionless
    density gradient for different $\alpha$. The SCAN functional is
    constructed based on 3 different limits of $\alpha$ corresponding
    to the different bonding regimes indicated.}
  \label{fig:exchange_enhancement}
\end{figure}

To illustrate the degree to which SCAN differs from the first two
rungs of Jacob's ladder, in Fig. \ref{fig:exchange_enhancement} we
plot the non-spin-polarized exchange enhancement factor versus the
reduced density gradient $s=|\nabla\rho|/[2(3\pi^2)^{1/3}\rho^{4/3}]$,
for different $\alpha$. SCAN distinguishes between three different
bonding regimes: metallic ($\alpha=1$), covalent ($\alpha=0$), and
weak ($\alpha\rightarrow\infty$). Interpolation between these limiting
values is used for other values of $\alpha$. Just as PBE is built
around LDA, SCAN is built around PBE. Therefore, just as the
$E_x^{PBE} \rightarrow E_{x}^{LDA}$ for $s \rightarrow 0$, in the
metallic regime ($\alpha=1$), $E_x^{SCAN} \rightarrow E_{x}^{PBE}
\rightarrow E_{x}^{LDA}$ for the same limit. However, for
not-so-slowly varying densities and/or values of $\alpha$ different
from unity, SCAN shows significantly different behavior than PBE and
LDA.

A limited amount of benchmarking has been performed on SCAN with
respect to solid-state materials. Sun \textit{et al.} computed the
lattice constant mean average error for 20 simple elemental and binary
solids and found values of 0.081, 0.059, and 0.016 \AA\ for LDA, PBE,
and SCAN, respectively \cite{sun_strongly_2015}. Tran \textit{et al.}
benchmarked a plethora of functionals at the LDA, GGA, meta-GGA, and
hybrid levels of theory \cite{tran_rungs_2016}. In addition to
computing the lattice parameter, cohesive energy, and bulk modulus for
44 strongly-bound elemental and binary solids (e.g. Pd, LiF), the
lattice parameter and cohesive/binding energy were computed for 5
weakly-bound solids (e.g. Ne, graphite). For the strongly-bound
solids, SCAN was found to have the lowest mean absolute relative
errors for all the computed properties as compared to the
commonly-used LDA, PBE, revPBE, PBEsol, BLYP, TPSS, and PBE0
functionals. The performance of SCAN for the weakly-bound solids was
less impressive on an absolute scale (mean absolute relative error in
the cohesive/binding energy of over 55\%), but again here SCAN
out-performed the other common functionals. Several (though not all
\cite{charles_assessing_2016}) other recent studies on small numbers
of systems are suggestive that SCAN is a significant improvement over
LDA and GGA for solid-state materials
\cite{sun_accurate_2016,kitchaev_energetics_2016,tian_exchange-correlation_2016,kylanpaa_accuracy_2017,paul_accuracy_2017,yao_plane-wave_2017,zhang_comparative_2017,hinuma_comparison_2017,bokdam_assessing_2017}.

In the present work, we perform an extensive benchmark of SCAN for a
diverse set of over 1,000 solids. We focus on the formation energy,
which is a central and widely-used quantity describing the
thermodynamic stability of solid compounds. We also present results
for crystal volume, magnetism, and band gap. In all cases, we compare
SCAN to the GGA (PBE) level of theory. PBE is chosen due to the
connection to SCAN as well as its prevalence and popularity
\cite{dftpoll}. We find SCAN performs remarkably well for
strongly-bound compounds, with a decrease in the formation energy mean
average error of around 50\% to 110 meV/atom relative to PBE. For less
strongly-bound compounds, i.e. intermetallic compounds, SCAN shows no
improvement compared to PBE; in fact, it provides moderately worse (by
around 20\%) formation energy predictions. The distinct exchange
behavior of SCAN in the covalent, metallic, and weak bonding regimes
is responsible for such trends. SCAN shows significant improvement in
predicted crystal structures. In particular, we find a mean average
volume error 40\% lower than than of PBE. SCAN provides moderately
improved band gap predictions compared to PBE, but it still has much
larger errors than fully nonlocal functionals such as hybrid
functionals and many-body perturbation theory approaches. Overall,
SCAN is a significant advance in describing strongly-bound compounds
at a modest increase in computational cost.

\section{Methodology}\label{sec:methodology}

\subsection{Compound formation energy}

We benchmark solid-state thermodynamics via the formation
energy \begin{equation} \Delta E_f=E - \sum_i x_i \mu_i.
  \label{eq:fe}
\end{equation}
Here $E$ is the total energy of a compound containing $x_i$ atoms of
element $i$, which has an elemental chemical potential of $\mu_i$, in
the formula unit. For example, for FeS$_2$, $\Delta E_f = E_{FeS_2} -
\mu_{Fe} - 2\mu_{S}$. In this work, all formation energies are
normalized to the number of atoms in the compound formula unit. We
assume the $PV$ term is small for the solid materials studied here,
i.e., $\Delta E_f \approx \Delta H_f$, where $\Delta H_f$ is the
formation enthalpy. Therefore, in the text we use $\Delta E_f$ and
$\Delta H_f$ interchangeably.

\subsection{Elemental reference states}

The elemental chemical potentials $\mu$ correspond to the energy per
atom of the pure element in a particular reference state. Here we
choose the elemental reference states to best match the experiments
since we compare to measured formation energies. These reference
states generally correspond to the stable phase at standard conditions
\cite{dinsdale_sgte_1991}, with a few exceptions. In the case of P,
$\alpha$ white P is the reference state. Since this phase has a
complicated structure with partial occupancy, we choose $\beta$ white
P as our reference state \cite{aykol_phosphorus_2017}. These two
phases have similar structural motifs. Similarly, in the case of B we
choose $\alpha$ rhombohedral B rather than $\beta$ rhombohedral B. For
elements with diatomic gases as the reference phases, we choose the
isolated diatomic molecule as our reference phase. We also consider
Xe-containing compounds, for which we choose the isolated Xe atom as
the reference state. The full list of elemental reference states is
given in the Supporting Information

\subsection{Compound selection}

We use the Open Quantum Materials Database (OQMD)
\cite{saal_materials_2013,kirklin_open_2015} to acquire the compound
and elemental crystal structures, as well as the tabulated
experimental formation energies. The experimental formation energies
come from two sources: the Scientific Group Thermodata Europe Solid
Substance (SSUB) database \cite{sgte_thermodynamic_1999} and the
thermodynamic database of the Thermal Processing Technology Center at
Illinois Institute of Technology (IIT) \cite{nash_thermodynamic_2013}.
Unlike SSUB, the IIT database focuses on intermetallic compounds.

We find the set of compounds in the OQMD for which the following
criteria are satisfied:
\begin{enumerate}
\item Compound does not contain Br or Hg
\item There is an experimental formation energy reported for the
  corresponding composition
\item The compound is reported in the Inorganic Crystal Structure
  Database (ICSD) \cite{bergerhoff_inorganic_1983,belsky_new_2002}
\end{enumerate}
Criterion 1 is chosen since Br and Hg are liquids at standard
conditions, which are more difficult to model. We note that the
experimental formation energies are tabulated by composition, rather
than by the precise structure, so criterion 2 does not always uniquely
identify a \textit{single} compound with the composition in the case
of polymorphism. Criterion 3 is chosen to pick out the structure most
likely to correspond to the experimentally measured formation energy.
In the case of multiple distinct structures present in ICSD at the
composition, we choose the lowest-energy compound (based on
calculations in the OQMD). This ensures only a single compound is
associated with each composition with a measured formation energy.
1,793 unique compounds in the OQMD satisfy these criteria \footnote{As
  of September 2017}.

To reduce computational cost, we choose the compounds whose primitive
unit cells contain no more than 10 atoms. This corresponds to 1,000
compounds. The distribution of compounds in terms of number of atoms
in the primitive cell is included in the Supporting Information. 912
of the compounds are binary, 87 are ternary, and 1 (CaMg(CO$_3$)$_2$)
is quaternary. 55 of the 1,000 selected compounds are ignored since
either (1) there is a significant discrepancy in different reported
experimental formation energies, (2) the magnitude of the experimental
formation energy is less than 50 meV/atom, and/or (3) the DFT
calculation for the compound or any constituent element failed to
converge for one or both of the xc functionals. This leaves 945 total
compounds. The rationale for excluding these compounds is discussed in
more detail in the Supporting Information. We also include the full
list of selected compounds, in addition to information on a few
exceptions for compound selection, experimental formation energy
values, and experimental volume values.



\section{Computational Details}\label{sec:compdetails}

DFT calculations are performed using the projector augmented wave
(PAW) method \cite{blochl_projector_1994,kresse_ultrasoft_1999} with a
600 eV plane wave cutoff using the Vienna \textit{ab initio} software
package (\textsc{vasp})
\cite{kresse_ab_1994,kresse_ab_1993,kresse_efficient_1996,kresse_efficiency_1996}.
We employ the recommended \textsc{vasp} 5.2 PBE PAW potentials for all
calculations (SCAN and PBE) \cite{vasp_paw} since SCAN PAW potentials
do not currently exist. This represents an approximation (albeit a
widely-used one) as the recent work of Yao and Kanai showed that the
use of PBE potentials for SCAN calculations can lead to differences in
certain cases \cite{yao_plane-wave_2017}. Uniform $\Gamma$-centered
Monkhorst-Pack $k$-point meshes \cite{monkhorst_special_1976} with
$k$-point density of at approximately 700 $k$-points per \AA$^{-3}$ or
greater are chosen. The average number of $k$-points times the number
of atoms in the unit cell, another metric of $k$-point density, is
approximately 11,300. 1st-order Methfessel-Paxton smearing
\cite{methfessel_high-precision_1989} of 0.2 eV is employed for
structural relaxations, while total energy calculations use the
tetrahedron method with Bl\"{o}chl corrections
\cite{blochl_improved_1994}. The energy and ionic forces are converged
to 10$^{-6}$ eV and 10$^{-3}$ eV/\AA, respectively. Spin-polarized
calculations with ferromagnetic initialization of 3.5 $\mu_B$ per
magnetic site are employed for compounds containing Sc--Cu, Y--Ag,
Lu--Au, La--Yb, and Ac--No; such initialization is also employed for
O$_2$ to properly capture the triplet ground state. For elements with
gaseous reference states, the isolated atom/molecule is computed using
a face-centered-cubic cell with 15 \AA\ conventional cell lattice
parameter and 50 meV Gaussian smearing. We note that our calculations
are performed using tighter convergence parameters than the existing
PBE-based calculations in the OQMD.

\section{Results and Discussion}\label{sec:results}

\subsection{Formation energy}

\begin{figure}[htbp]
  \begin{center}
    \includegraphics[width=1.0\linewidth]{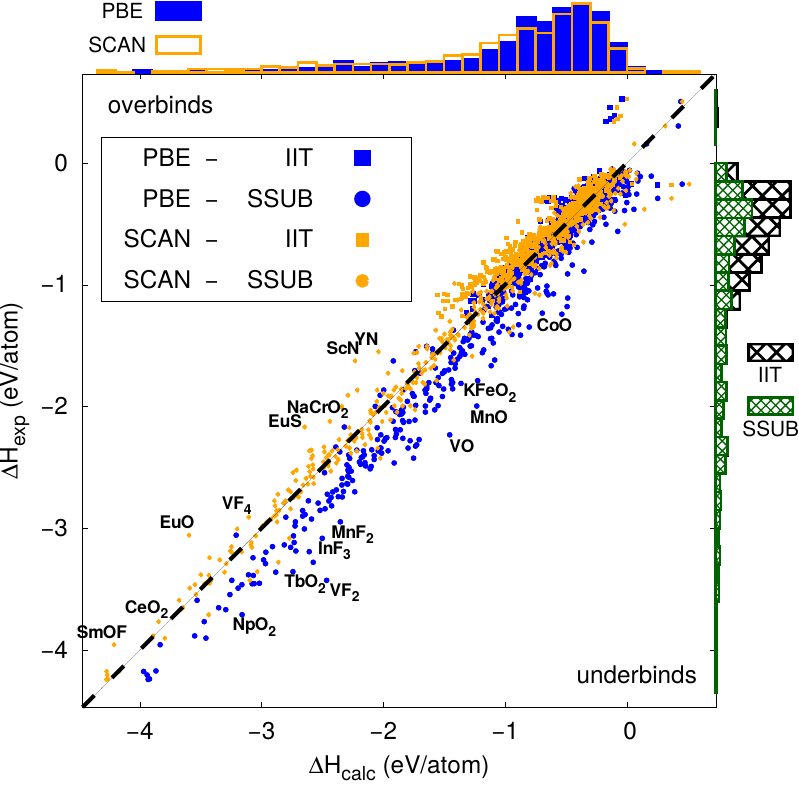}
  \end{center}
  \caption{Comparison of calculated and experimental experimental
    formation energy for the 945 compounds for PBE and SCAN. Multiple
    points for the same compound and functional correspond to
    different sources of experimental formation energy. The dashed
    diagonal line corresponds to the $\Delta H_{calc}=\Delta H_{exp}$
    line of perfect agreement. Each bar chart on the axes corresponds
    to a histogram, which is stacked in the case of the experimental
    formation energy.}
  \label{fig:fe}
\end{figure}

The comparison of computed and experimental formation energies is
shown in Fig. \ref{fig:fe}, which represents the primary result of
this work. The most striking trend is the significant improvement of
SCAN over PBE for the predicted $\Delta H$ for compounds with a large,
negative formation energies. We refer to these compounds as
strongly-bound compounds. This trend is seen most dramatically for
values of $\Delta H_{exp}$ of around $-1$ to $-4$ eV/atom. Here the
magnitude of $\Delta H_{calc}$ for PBE is significantly lower than
that of experiment in a systematic fashion, corresponding to an
underbinding of the compound with respect to the elements. In stark
contrast, no clear systematic underbinding or overbinding of $\Delta
H_{calc}$ for SCAN is apparent. We note that the SCAN values for
strongly-bound compounds still have deviations (both positive and
negative) from the $\Delta H_{calc}=\Delta H_{exp}$ line. A
quantitative analysis of the errors, for both functionals, will be
presented further below. For compounds with a smaller magnitude of
$\Delta H_{exp}$, which we call weakly-bound compounds, the
differences in the accuracy of the prediction for PBE and SCAN are
less obvious. As shown in the histogram of $\Delta H_{exp}$ in Fig.
\ref{fig:fe}, these weakly-bound compounds represent the majority of
the experimental data. This stems in part from the focus of the IIT
database on intermetallic compounds, which generally have
low-magnitude $\Delta H$.

\begin{figure*}[tb]
  \begin{center}
    \includegraphics[width=1.0\linewidth]{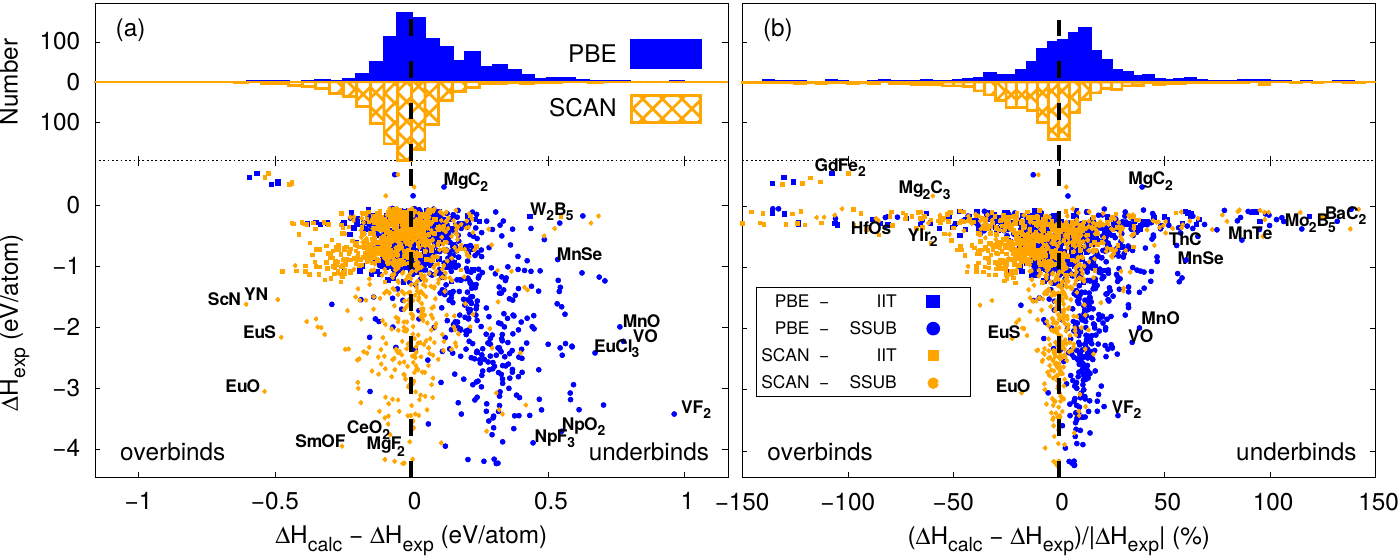}
  \end{center}
  \caption{(a) Absolute and (b) relative errors of the calculated
    formation energies plotted against the experimental formation
    energies. The dashed vertical lines correspond to the $\Delta
    H_{calc}=\Delta H_{exp}$ line of perfect agreement. Histograms of
    the errors are shown at the top. For the relative errors, the
    range is limited to $\pm 150\%$.}
  \label{fig:fe_errors}
\end{figure*}

In order to more easily see the difference in accuracy between $\Delta
H_{calc}$ of PBE and SCAN for the full range of $\Delta H_{exp}$, in
Fig. \ref{fig:fe_errors}(a) we plot the error $\Delta H_{calc}-\Delta
H_{exp}$ as a function of $\Delta H_{exp}$. Histograms of the errors
are also included. Here PBE's systematic underbinding of $\Delta H$
for strongly-bound compounds can again be observed in the region of
$\Delta H_{exp}$ of around $-1$ to $-4$ eV/atom. In addition, the
errors appear to increase with increasing magnitude of $\Delta
H_{exp}$. In other words, the more negative $\Delta H_{exp}$ is, the
more PBE underbinds, which has been previously observed
\cite{kirklin_open_2015}. In the histogram of error values for PBE,
this systematic underbinding leads to a longer tail of positive error
values, resulting in a distribution that appears to be centered around
a positive value rather than zero. No such strong systematic
underbinding or overbinding for strongly-bound compounds can be
observed for SCAN. As such, the distribution of the errors for SCAN
appears much better (though not perfectly) centered around zero.

In Fig. \ref{fig:fe_errors}(a), one can also observe a moderate
\textit{overbinding} trend of SCAN for the weakly-bound compounds. For
example, for $\Delta H_{exp}$ of around $-1$ to $-0.5$ eV/atom, many
more of the SCAN error values are negative (corresponding to
overbinding) than positive. In contrast, such an overbinding tendency
for weakly-bound compounds is not observed for PBE. The moderate,
systematic overbinding of SCAN for weakly-bound compounds contributes
to a slightly negative (overbinding) center-of-mass of the error
distribution.

These two trends -- SCAN's lack of systematic underbinding for
strongly-bound compounds and its mild overbinding of weakly-bound
compounds -- are our primary findings in terms of compound formation
energy. The same trends can be observed in the \textit{relative}
errors $(\Delta H_{calc}-\Delta H_{exp})/|\Delta H_{exp}|$, shown in
Fig. \ref{fig:fe_errors}(b). Consistent with the absolute error data,
PBE exhibits a roughly constant positive average relative error (on
the order of +10\%) for strongly-bound compounds, contributing to a
skewing of the relative error histogram to positive (underbinding)
values. In contrast, no clear systematic underbinding or overbinding
for strongly-bound compounds is found for SCAN. For the weakly-bound
compounds, the relative errors can blow up for compounds with small
$\Delta H_{exp}$ due to the $|\Delta H_{exp}|$ in the denominator of
the relative error. For example, a relative error of over 240\% in
magnitude is found for TaCo$_2$, which has $\Delta H_{exp}$ of only
$-0.065$ eV/atom. Therefore, the relative error axis is truncated in
Fig. \ref{fig:fe_errors}(b) for clarity. One can again observe SCAN's
moderate, systematic overbinding of weakly-bound compounds with
$\Delta H_{exp}$ of around $-0.5$ to $-1$ eV/atom, contributing to a
clear shoulder in the relative error histogram at around $-20$\%.

\begin{figure}[htbp]
  \begin{center}
    \includegraphics[width=1.0\linewidth]{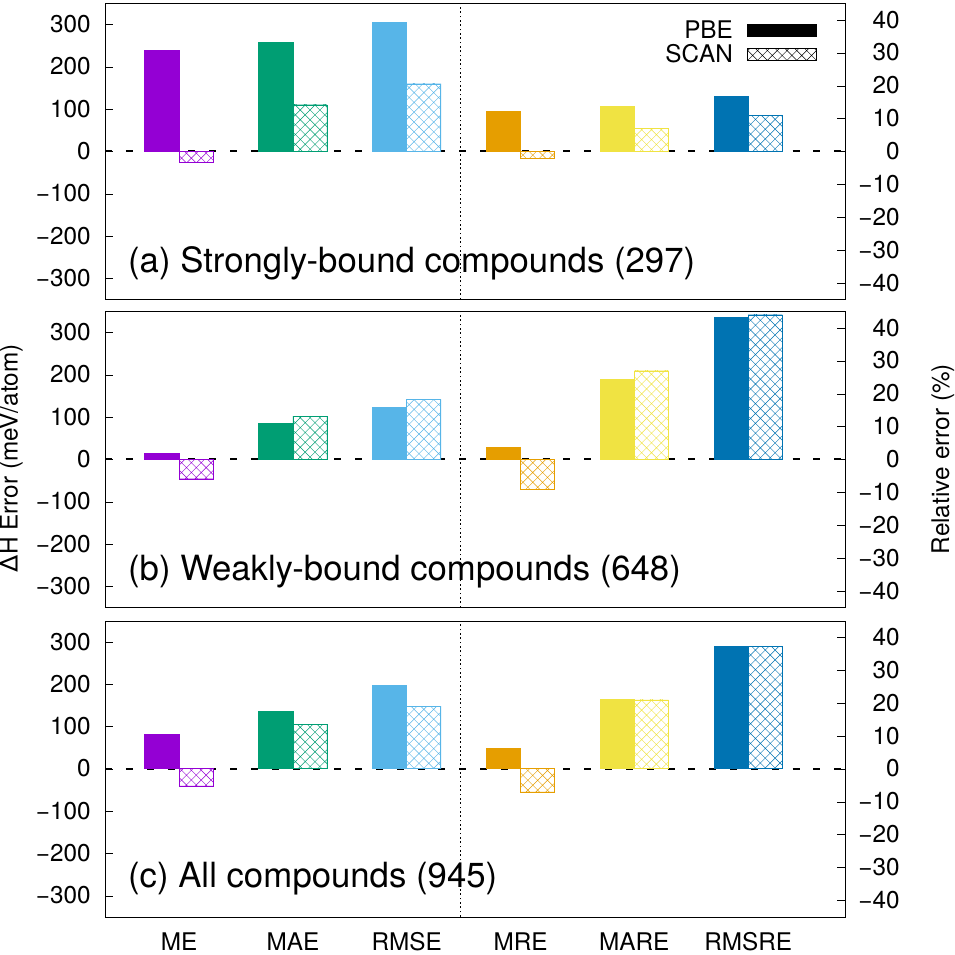}
  \end{center}
  \caption{Comparison of absolute and relative errors for PBE and SCAN
    for (a) 297 strongly-bound compounds, (b) 648 weakly-bound
    compounds, and (c) all 945 compounds. Absolute (relative) errors
    are plotted with respect to the left (right) axis.}
  \label{fig:stats}
\end{figure}

In order to separately analyze the formation energy errors for
strongly- and weakly-bound compounds, we partition the total 945
compounds into these two groups based on $\Delta H_{exp}$. We choose
to define strongly-bound compound as any compound with $\Delta
H_{exp}<-1$ eV/atom; the rest are weakly-bound compounds. The critical
value of precisely $-1$ eV/atom is somewhat arbitrary and is chosen
based on visual inspection of the different regions of data in Fig.
\ref{fig:fe}. However, we find no change in the conclusions discussed
below by slightly varying this value. In addition, performing an
analogous analysis based on partitioning the compound set based on
constituent elements rather than $\Delta H_{exp}$, discussed in the
Supporting Information, also leads to the same conclusions.

Using this $\Delta H_{exp}$-based convention for compound set
partitioning, we quantify the formation energy errors for strongly-
and weakly-bound compounds in Figs. \ref{fig:stats}(a) and
\ref{fig:stats}(b), respectively. Mean error (ME), mean average error
(MAE), root-mean-square error (RMSE), mean relative error (MRE), mean
absolute relative error (MARE), and root-mean-square relative error
(RMSRE) are considered based on the error and relative error
expressions discussed above. For each subset of compounds, the
absolute and relative error metrics show the same qualitative trend.
In particular, by all error metrics SCAN out-performs PBE for the
strongly-bound compounds, whereas PBE out-performs SCAN for the
weakly-bound compounds. For strongly-bound compounds, SCAN has a ME of
only $-0.027$ eV/atom as compared to $+0.239$ eV/atom for PBE. In
other words, the distribution of errors for SCAN's description of
strongly-bound compounds is extremely well-centered around zero. This
enables much improved MAE and RMSE values for SCAN (0.110 and 0.159
eV/atom) as compared to PBE (0.259 and 0.305 eV/atom). These
correspond to very significant ($\approx 50\%$) decreases in error.
The maximum absolute error for SCAN ($-0.605$ eV/atom for ScN) is also
substantially decreased as compared to that of PBE ($0.963$ eV/atom
for VF$_2$). Similarly large improvements for SCAN are found in terms
of the relative errors, with 82\%, 49\%, and 34\% decreases in the
magnitudes of the MRE, MARE, and RMSRE values, respectively.

For weakly-bound compounds, we find the opposite qualitative trend.
PBE is moderately more accurate than SCAN for such compounds. The ME
for PBE is significantly smaller in magnitude ($0.015$ eV/atom) than
that of SCAN ($-0.046$ eV/atom). However, SCAN's increases in MAE and
RMSE are a more modest 20\% and 16\%, respectively, compared to the
PBE values. Similarly, for the relative errors, the increase in MARE
and RMSRE for SCAN compared to those of PBE are 10\% and 2\%,
respectively. Ultimately, we find SCAN is significantly better for
predicting the formation energy of strongly-bound compounds, while it
is moderately worse for weakly-bound compounds like intermetallics.

Finally, we briefly comment on the overall $\Delta H_{calc}$ errors
(for all 945 compounds). These errors, plotted in Fig.
\ref{fig:stats}(c), reflect the combination of (1) significant error
reduction for strongly-bound compounds with large $\Delta H_{exp}$
magnitude and (2) modest error increase for weakly-bound compounds
with small $\Delta H_{exp}$ magnitude. In our particular case, there
are 648 weakly-bound compounds and only 297 strongly-bound compounds.
In this case, SCAN achieves modest decreases in absolute errors with
decreases in MAE and RMSE of 24\% and 25\%, respectively, and
essentially no difference in terms of relative errors. We emphasize
that the overall formation energy errors in SCAN will be a strong
function of the fractions of strongly- and weakly-bound compounds
under consideration. Therefore, one should consider the two individual
formation energy error trends, rather than the overall trend for our
particular compound set, as the key result.




\subsection{Origin of formation energy trends}

In order to elucidate the origin of the distinct formation energy
trends for strongly- and weakly-bound compounds, we perform a detailed
analysis of the strongly-bound oxide CaO and three weakly-bound
intermetallic compounds HfOs, ScPt, and VPt$_2$. While for CaO SCAN
reduces the $\Delta H_{calc}$ error from 324 meV/atom in PBE down to
just 2 meV/atom, for the intermetallics SCAN increases the error
magnitudes by 181--261 meV/atom. We note that both LDA and the earlier
meta-GGA of Tao, Perdew, Staroverov, and Scuseria (TPSS)
\cite{tao_climbing_2003} exhibit similar $\Delta H_{calc}$ as PBE
(differences no larger than 19 meV/atom in magnitude) for these
intermetallic compounds, which suggests the overbinding of this class
of compounds is specific to SCAN.

\begin{figure}[htbp]
  \begin{center}
    \includegraphics[width=1.0\linewidth]{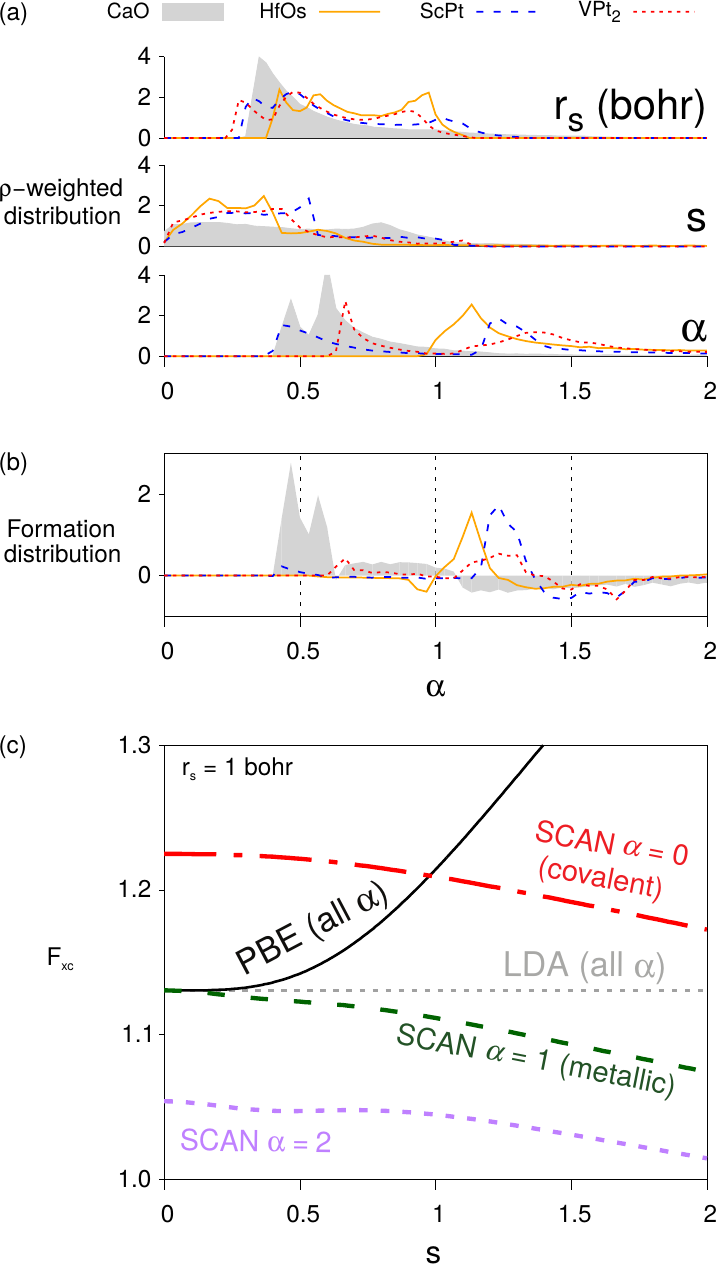}
  \end{center}
  \caption{(a) Density-weighted probability distributions of $r_s$,
    $s$, and $\alpha$ in real space for CaO and three intermetallic
    compounds. The distributions are normalized to unity. (b) The
    compound $\alpha$ formation distribution defined in analogy to
    the formation energy of Eq. \ref{eq:fe}. (c) Non-spin-polarized
    exchange-correlation enhancement factor at $r_s=1$ bohr for LDA,
    PBE, and SCAN at several relevant $\alpha$ values.}
  \label{fig:alpha}
\end{figure}

Figure \ref{fig:alpha}(a) contains normalized real-space
distributions, for each compound, of the three ingredients to
$\epsilon_{xc}$ for SCAN: $\rho$, $s$, and $\alpha$. The density is
parametrized by the Wigner-Seitz radius $r_s = (3/4\pi\rho)^{1/3}$ for
convenience. We weight the distributions by the density $\rho$ since
higher-density regions of space have a higher impact on $E_{xc}$ due
to the explicit factor of $\rho$ in Eq. \ref{eq:mgga_exc}. The results
are shown for SCAN calculations, though we show in the Supporting
Information that the distributions are similar for the PBE case.

The distribution of $\alpha$ shows that, unlike CaO, the intermetallic
compounds exhibit significant (even dominant in the case of HfOs)
probability of larger $\alpha$ in the range of 1--2. In addition, the
distributions illustrate the overall relevant ranges of parameter
space for the four compounds: $\sim$~0.25--1.25 for $r_s$, $\sim$~0--1
for $s$, and $\sim$~0.25--2 for $\alpha$. Including the elemental
reference states, whose $\alpha$ and $s$ distributions are shown in
the Supporting Information, one finds expanded ranges of $\sim$
0.25--1.75 for $r_s$, $\sim$~0--2 for $s$, and $\sim$~0.25--2 for
$\alpha$. We note that O$_2$ has a significantly broader distribution
of $s$ and $\alpha$, which is consistent with its distinct molecular
nature.

We construct the corresponding \textit{formation distribution} of
$\alpha$ by computing the difference between the $\alpha$ distribution
of the compound and the appropriate linear combination of the $\alpha$
distributions of the constituent elements, in analogy to Eq.
\ref{eq:fe}. This function, shown in Fig. \ref{fig:alpha}(b)
represents the change in the distribution of $\alpha$ that occurs upon
formation of the compound from the elements. The formation
distributions show appreciable rearrangement of $\alpha$ upon compound
formation. There are distinct behaviors for CaO and for the
intermetallics that explain the distinct formation energy behavior for
SCAN, which we now will discuss.

The CaO formation distribution exhibits a large positive peak in the
vicinity of $\alpha=0.5$, balanced by decreased probability of
$\alpha>1$. This peak stems from the filling of the O $p$ shell,
whereas the broad valley for larger $\alpha$ corresponds to the
depletion of O$_2$ states. To understand the impact on $E_{xc}$, in
Fig. \ref{fig:alpha}(c) we plot the non-spin-polarized
exchange-correlation enhancement factor $F_{xc} = E_{xc}/E_x^{LDA}$ as
a function of $s$ for LDA, PBE, and SCAN for several $\alpha$. As
compared to the exchange enhancement in Fig.
\ref{fig:exchange_enhancement} discussed previously, $F_{xc}$ adds the
density-dependent contribution from correlation. The values in Fig.
\ref{fig:alpha}(c) are plotted for $r_s$ of 1 bohr as a representative
example; $F_{xc}$ for other relevant values of $r_s$ (shown in the
Supporting Information) show the same qualitative behavior. The SCAN
$F_{xc}$ increases with decreasing $\alpha$. Therefore, the increased
probability for lower $\alpha$ upon formation of CaO leads to a
negative $E_{xc}$ contribution to $\Delta H_{calc}$ within SCAN since
$E_x^{LDA} = -(3/4\pi)(3\pi^2\rho)^{1/3}$ is a negative energy. In
contrast, PBE as a GGA has no dependence on $\alpha$, so it lacks this
negative contribution to $\Delta H_{calc}$. SCAN thus predicts a more
negative $\Delta H_{calc}$ than PBE, leading to much better agreement
of $\Delta H_{calc}$ with experiment, due to its behavior in the
$\alpha=0$ (covalent bonding) regime as compared to that of larger
$\alpha$. We attribute the improvement to the exchange energy in
particular since this is the largest energetic term. We note that a
more rigorous analysis for CaO should consider the spin-dependent
$F_{xc}$ since O$_2$ is in a triplet state, but we expect the same
qualitative trend given the differences between $F_{xc}$ for the
non-spin-polarized and fully spin-polarized cases shown in Ref.
\citenum{sun_strongly_2015}.


For the intermetallics, the $\alpha$ formation distribution shows the
most significant rearrangement in the regime of $\sim$~1--1.7. In
particular, for each compound there is a positive contribution for
$\alpha$ of 1.1--1.3 and a negative contribution for larger $\alpha$
of $\sim$~1.4--1.7. In other words, intermetallic compound formation
leads to smaller $\alpha$ in the $\alpha>1$ regime. This rearrangement
stems from the decreased weak bonding in the compounds as compared to
elements like Hf, Os, and Pt. As shown in Fig. \ref{fig:alpha}(c), for
$\alpha>1$, smaller $\alpha$ again leads to increased $F_{xc}$ and
thus a more negative $E_{xc}$. SCAN thus predicts a more negative
$\Delta H_{calc}$ than PBE, in this case leading to moderately worse
agreement with experiment, due to its behavior in the
$\alpha\rightarrow\infty$ (weak bonding) regime. We again attribute
the exchange energy in particular since it is the largest energetic
term. This finding is consistent with the very different $\Delta
H_{calc}$ error for the intermetallic compounds found within TPSS,
which exhibits quite distinct $F_x$ behavior for $\alpha>1$
\cite{tao_climbing_2003}.

\subsection{Elemental chemical potential fitting}

\begin{figure}[htbp]
  \begin{center}
    \includegraphics[width=1.0\linewidth]{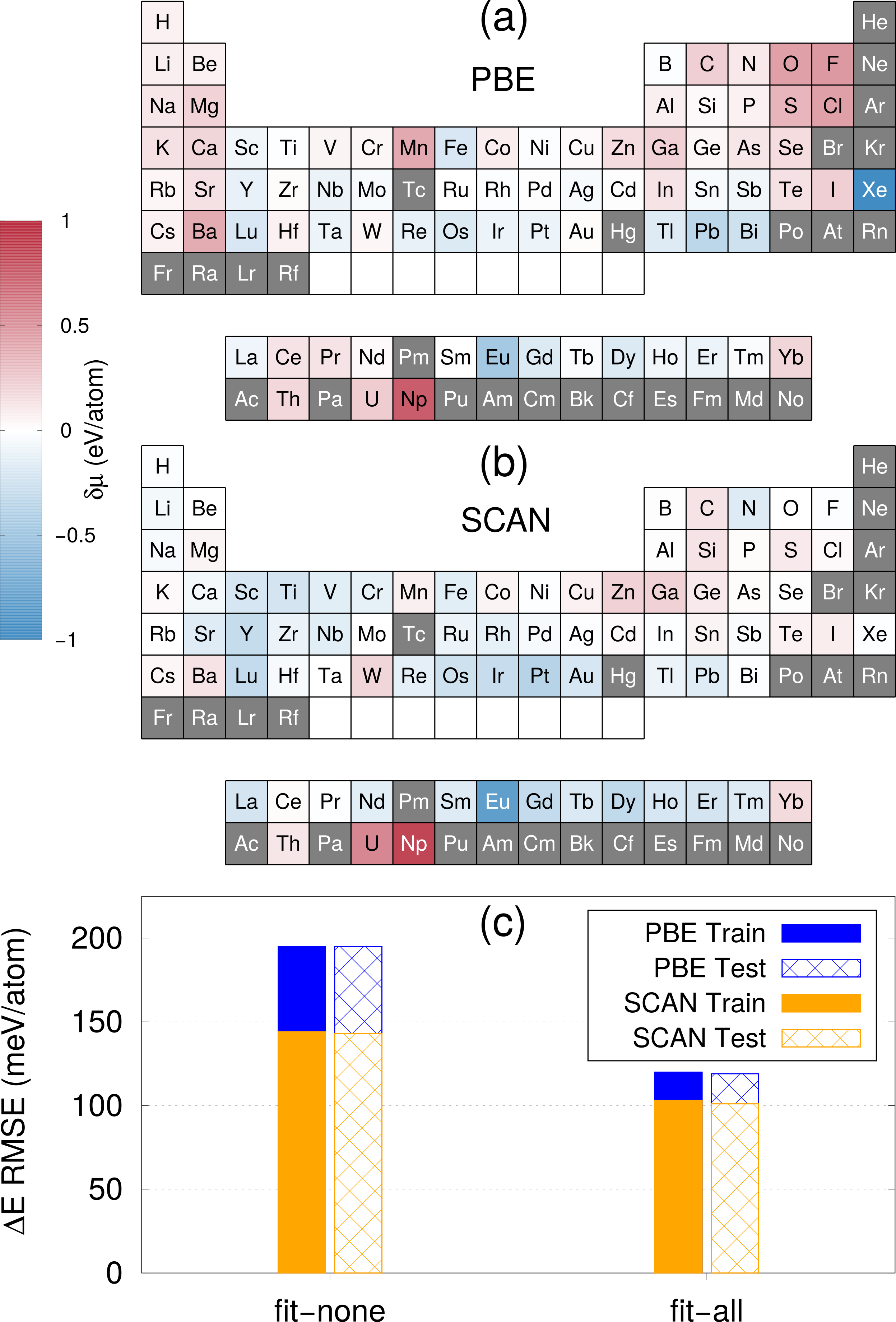}
  \end{center}
  \caption{Elemental chemical potential corrections $\delta\mu$
    obtained via fitting for (a) PBE and (b) SCAN. Grey-colored
    squares correspond to elements not considered in the compound set.
    The training and testing $\Delta E$ RMSE from 9-fold
    cross-validation are shown in panel (c) for the cases of fitting
    $\mu$ for no (``fit-none'') and all (``fit-all'') elements.}
  \label{fig:mu_fitting}
\end{figure}

One approach to improve the quality of the predicted $\Delta H$, at
the cost of adding empiricism, is to fit the elemental chemical
potential $\mu$ for one or more elements
\cite{wang_oxidation_2006,stevanovic_correcting_2012,grindy_approaching_2013,kirklin_open_2015}.
Here we perform a simultaneous least-squares fitting of $\mu$ for all
78 periodic table elements (``fit-all'') contained within our set of
945 compounds, which minimizes the RMSE of $\Delta H$. The resulting
corrections to the DFT-calculated $\mu$ (which we call $\delta\mu$)
are plotted in Figs. \ref{fig:mu_fitting}(a) and
\ref{fig:mu_fitting}(b) for PBE and SCAN, respectively. For PBE,
significant \textit{positive} corrections are found for
electronegative elements such as O, S, F, and Cl. These corrections
are consistent with PBE's tendency to underbind the strongly-bound
compounds, which typically contain such elements. In contrast, since
SCAN does not suffer from an appreciable systematic underbinding or
overbinding of the strongly-bound compounds, there are no large
corrections to the electronegative elements for SCAN. Since SCAN
moderately overbinds intermetallic compounds, there are many more
\textit{negative} $\delta\mu$ for metallic elements for SCAN as
compared to those of PBE. One can observe this trend for many alkali,
alkaline earth, transition, and lanthanide metals.

In order to assess the possibility of overfitting, we perform 9-fold
cross-validation and separately analyze the training and testing
errors. Figure \ref{fig:mu_fitting}(c) illustrates the RMSE $\Delta H$
errors for PBE and SCAN for the fit-all case and that of no fitting
(``fit-none''). Fitting reduces the error from 195 to 120 meV/atom for
PBE and 144 to 104 meV/atom for SCAN. Since the training and testing
errors are nearly identical, this indicates there is no overfitting in
predicting $\Delta H$ with the set of fit $\mu$. We note that the
corresponding testing MAE are 83 and 72 meV/atom for PBE and SCAN,
respectively. The full set of fitted $\mu$ values are included in the
Supporting Information.

\subsection{Volumes}

\begin{figure}[htbp]
  \begin{center}
    \includegraphics[width=1.0\linewidth]{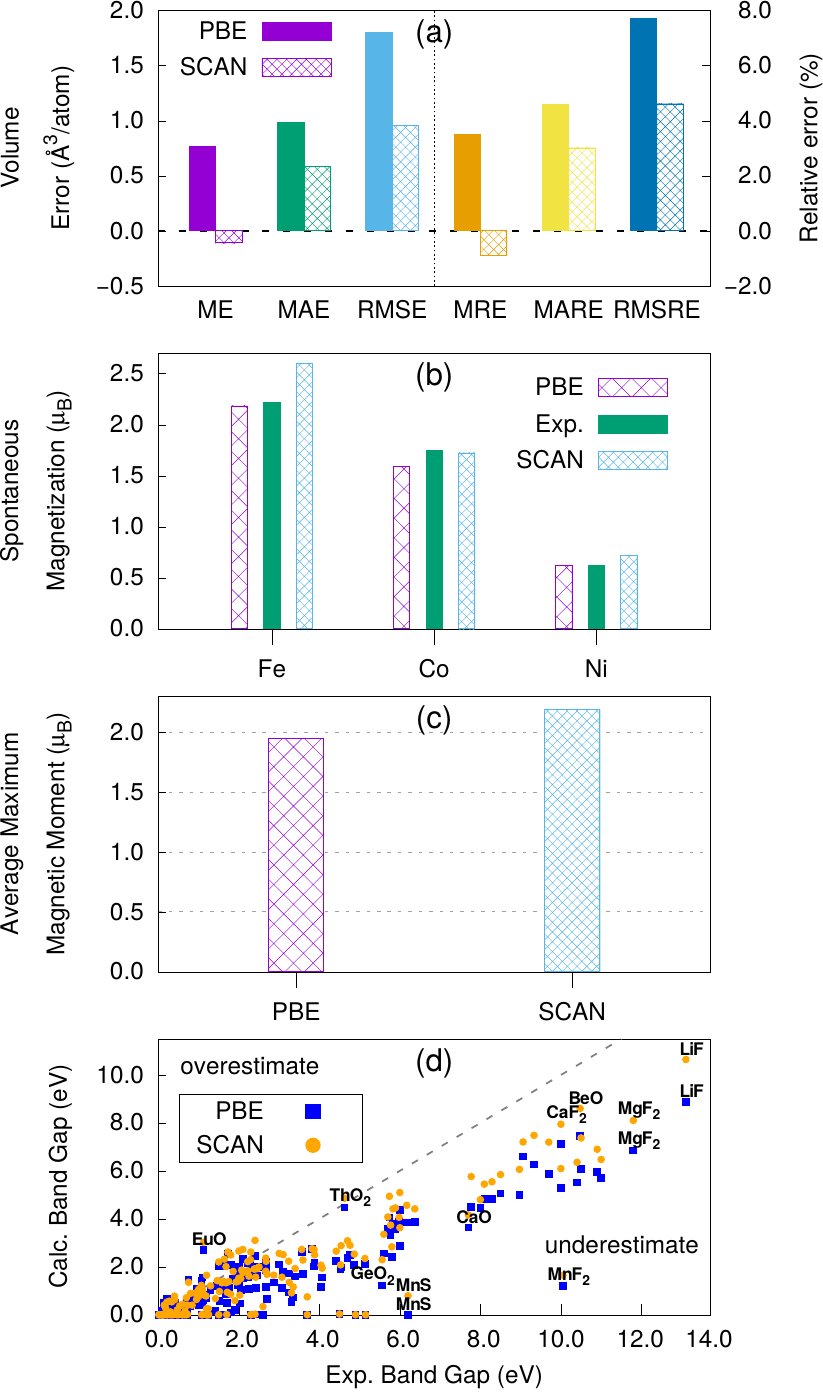}
  \end{center}
  \caption{Comparison of predicted structural, magnetic, and
    electronic properties for PBE and SCAN, with experimental values
    shown when available. (a) Absolute and relative errors in compound
    volume. (b) Spontaneous magnetization for Fe, Co, and Ni. (c)
    Average maximum local magnetic moment for magnetic compounds. (d)
    Calculated and experimental electronic band gap, with the dashed
    line corresponding to perfect agreement.}
  \label{fig:other_properties}
\end{figure}

In order to assess the accuracy of the crystal structures, we compare
the computed relaxed volume per atom to experimental values taken from
the ICSD. Both the compounds and solid elements are considered. Figure
\ref{fig:other_properties}(a), which shows the absolute and relative
errors for PBE and SCAN, demonstrates that SCAN achieves a significant
improvement in the predicted volumes. While PBE on average
overestimates the volume by 0.77 \AA$^3$/atom, SCAN only
underestimates it by 0.11 \AA$^3$/atom. The MAE for SCAN is 0.58
\AA$^3$/atom, a 41\% decrease from the corresponding PBE value. The
relative error magnitudes for SCAN are similarly smaller than those of
PBE.

A complete plot of the predicted and experimental volumes, shown in
the Supporting Information, illustrates that the SCAN's improved
prediction of volume is particularly significant for layered
materials. For example, SCAN predicts a volume of 35.6 \AA$^3$/atom
for the layered material MgI$_2$ with experimental volume of 34.3
\AA$^3$/atom, whereas PBE significantly overestimates the volume with
a value of 40.6 \AA$^3$/atom. Such improved volume prediction for
layered materials is consistent with the improved treatment of
(intermediate) van der Waals interaction in SCAN
\cite{sun_strongly_2015}. We find similar behavior for one-dimensional
materials like BeI$_2$ and AgCN.

Unlike the behavior for $\Delta H$, the predicted volume behavior is
less different between strongly- and weakly-bound compounds. The same
qualitative trend, smaller errors for SCAN than PBE, is found for both
sets of compounds by all error metrics considered. In addition, for
SCAN the quantitative accuracy of the predicted volumes is essentially
the same for the two sets of compounds. The difference in volume MAE
between the two subsets of compounds is only 0.02 \AA$^3$/atom for
SCAN, with the slightly larger errors for strongly-bound compounds.
For PBE, the difference is larger (0.50 \AA$^3$/atom), also with the
larger errors for strongly-bound compounds. This is consistent with
the underbinding trend of $\Delta H$ for strongly-bound compounds and
suggests that PBE's treatment of the compound (as opposed to only the
elements) contributes to the underbinding of $\Delta H$. Additionally,
the larger errors for layered materials, which are mainly
strongly-bound compounds (in the sense of large $\Delta H_{exp}$) also
contribute to the worsened volume predictions for strongly-bound
compounds in PBE. For example, layered ZrCl$_2$ with $\Delta H_{exp}$
of -1.73 eV/atom exhibits a 44\% volume error within PBE (4\% in
SCAN). Overall, SCAN shows a significant improvement over PBE for
prediction of crystal volumes with 57\% and 33\% MAE reductions for
strongly- and weakly-bound compounds, respectively.

\subsection{Magnetism}

Next, we explore the predicted magnetic properties. The spontaneous
magnetization of the elemental metals Fe, Co, and Ni for PBE and SCAN
as compared to experiment \cite{crangle_magnetization_1971} are shown
in Fig. \ref{fig:other_properties}(b). In all cases, SCAN predicts a
moderately larger magnetization than that of PBE. The enhancement in
magnetization is 0.42 $\mu_B$ (19\%), 0.13 $\mu_B$ (8\%), and 0.1
$\mu_B$ (14\%) for Fe, Co, and Ni, respectively. For Fe and Ni, this
leads to worsened comparison to experiment. For example, SCAN
overestimates the magnetization of Fe by 17\%, while PBE
underestimates it by only 2\%. In contrast, SCAN's magnetization for
Co (1.72 $\mu_B$) is closer to the experimental value (1.75 $\mu_B$)
than that of PBE (1.59 $\mu_B$). Such results suggest a tendency of
SCAN to moderately overestimate the magnetism in itinerant
(ferro)magnets in some cases.

We also compare the predicted local magnetic moments of PBE and SCAN
in all of the magnetic systems considered. We choose to consider a
system magnetic if any local magnetic moment is greater than 0.1
$\mu_B$ in magnitude. Figure \ref{fig:other_properties}(c) shows the
maximum local magnetic moment, averaged over all the magnetic systems.
While the overall magnitude of this quantity (around 2 $\mu_B$) is not
important as it is dependent on the particular set of elements and
compounds studied in this work, we comment on the difference in the
values between PBE and SCAN. Consistent with the behavior for the
elemental ferromagnets, here SCAN again shows a moderate magnetic
enhancement. In particular, the average maximum magnetic moment within
SCAN is 12\% larger than that found within PBE. A complete plot of the
magnetic moments for all the magnetic compounds, which illustrates
this trend, is included in the Supporting Information. This plot also
indicates there are certain compounds predicted to be non-magnetic
within PBE for which SCAN predicts magnetism (e.g., FeTe$_2$ and
FeCl$_2$).

\subsection{Band gaps}

Finally, we consider the performance of SCAN for electronic band gap
prediction. Semilocal approximations to $E_{xc}$ like LDA and PBE are
well known to underestimate band gaps \cite{perdew_density_1985}.
Although SCAN is not specifically designed to address this band gap
problem, it is interesting to evaluate its accuracy for predicting
band gaps as compared to PBE, especially since SCAN in principle
contains some nonlocality via $\tau$. Details of the extraction of
experimental band gap values \cite{strehlow_compilation_1973} for
comparison are discussed in the Supporting Information.

Figure \ref{fig:other_properties}(d) compares computed band gaps to
experimental values. Nearly all the points lie below the dashed line
of perfect agreement, which indicates that SCAN like PBE suffers from
a systematic underestimation of electronic band gap. However, the SCAN
points are usually in better agreement with experiment. As one
example, the SCAN band gap for GaN is 2.2 eV as compared to 3.2 eV in
experiment, whereas the PBE gap is 1.7 eV. Additionally, it appears
the improvement in band gap prediction for SCAN as compared to PBE
becomes more significant on an absolute scale as the magnitude of the
band gap increases. For example, for LiF (with a very large
experimental gap of 13.1 eV), the SCAN band gap is 1.8 eV larger than
that of PBE, though still underestimating the experimental value. The
band gap enhancement of SCAN compared to PBE is not solely a result of
structural relaxation. For example, SCAN exhibits a 0.3 eV enhancement
of band gap of GaN with respect to PBE using the experimental
structure, as compared to 0.5 eV using relaxed structures. For LiF,
the corresponding band gap enhancement is 1.1 eV and 1.8 eV without
and with structural relaxations, respectively.

Overall, we find a band gap MAE of 1.2 eV for SCAN as compared to 1.5
eV for PBE. This indicates that SCAN provides a modest improvement in
band gap prediction as compared to PBE, though the band gaps still
significantly underestimate experimental values. We note that fully
nonlocal approaches to band gap prediction like many-body perturbation
theory in the $GW$ approximation
\cite{hedin_new_1965,hybertsen_first-principles_1985,hybertsen_electron_1986,faleev_all-electron_2004,vanschilfgaarde_quasiparticle_2006}
as well as hybrid functionals like HSE \cite{heyd_hybrid_2003} perform
significantly better at band gap prediction \cite{tran_accurate_2009}.
For example, previous work has shown band gap MAE of 0.6 eV for HSE
and 0.5 eV for $GW,$ considering the 15 compounds in Ref.
\citenum{tran_accurate_2009} for which self-consistent $GW$ and HSE
values are given. Another earlier work found a band gap MAE of 0.3 eV
for HSE \cite{heyd_energy_2005}.

\section{Conclusions}\label{sec:conclusions}

In summary, an extensive benchmark of the new SCAN meta-GGA for a
diverse set of approximately 1,000 inorganic crystals is performed and
compared to the GGA level of theory (PBE). Unlike PBE, SCAN does not
exhibit a substantial, systematic underbinding of strongly-bound
compounds with respect to the elements, due to enhanced exchange
interaction in the covalent bonding regime. This leads SCAN to
significantly out-perform PBE for formation energies of such
compounds, with a decrease in MAE of around 50\% to 110 meV/atom. In
contrast, due to distinct exchange behavior in the weak bonding
regime, SCAN performs moderately worse than PBE for weakly-bound
compounds like intermetallics, for which the formation energy MAE
increases by 20\% to 102 meV/atom. The formation energy errors can be
further reduced by fitting the elemental chemical potentials. SCAN
shows significant improvement in volume prediction, with a 41\%
decrease in MAE with respect to PBE to 0.58 \AA$^3$/atom. A moderate
magnetic enhancement is found using SCAN as compared to PBE, with a
12\% increase in the average maximum magnetic moment. SCAN
significantly underestimates experimental band gaps, though there are
moderate improvements (20\% decrease in MAE) as compared to PBE.
Overall, SCAN represents a significant improvement in accuracy for
strongly-bound compounds as compared to PBE.

\begin{acknowledgments}
We thank Georg Kresse (Univ. of Vienna) for useful discussions. We
acknowledge support from the U.S. Department of Energy under Contract
DE-SC0015106. Computational resources were provided by the Quest high
performance computing facility at Northwestern University and the
National Energy Research Scientific Computing Center (U.S. Department
of Energy Contract DE-AC02-05CH11231).
\end{acknowledgments}

\bibliography{scan_thermodynamics}

\end{document}